\documentclass[%
 reprint,
 amsmath,amssymb,
 aps,
 pra,
]{revtex4-2}

\usepackage{graphicx}
\usepackage{dcolumn}
\usepackage{bm}
\usepackage{amsfonts}
\usepackage{amsmath}
\usepackage{float}
\usepackage{caption}
\usepackage{color}
\usepackage{comment}
\usepackage{soul}

\usepackage{graphicx,subcaption}

\begin{document}

\preprint{APS/123-QED}

\title{Spring-loaded: Engineering Dispersion Relations Using Complex Couplings}

\author{R.~G. Edge, S.~A.~R. Horsley, T.~A. Starkey, and G.~J. Chaplain}

\affiliation{Centre for Metamaterial Research and Innovation, Department of Physics and Astronomy, University of Exeter, Exeter EX4 4QL, United Kingdom}

\date{\today}

\begin{abstract}
     Motivated by recent theoretical and experimental interest in metamaterials comprising non-local coupling terms, we present an analytic framework to realise materials with arbitrary complex dispersion relations. Building on the inverse design method proposed by Kazemi \textit{et al}. [Phys. Rev. Lett. 131, 176101 (2023)], which demonstrated arbitrary real dispersion relations with real non-local coupling terms, we show that using \textit{complex} non-local couplings not only can band structures be drawn, but that arbitrary attenuation of the supported modes can be achieved by tuning their trajectory in the complex frequency plane. This is demonstrated using a canonical mass-spring lattice with higher-order spatial connections; we consider the energy velocity in such systems and show that competing loss and gain of the complex interactions can produce exotic wave dispersion. 

\end{abstract}

\maketitle
\section{Introduction}

\begin{figure*}
    \centering
    \includegraphics[width=0.95\linewidth]{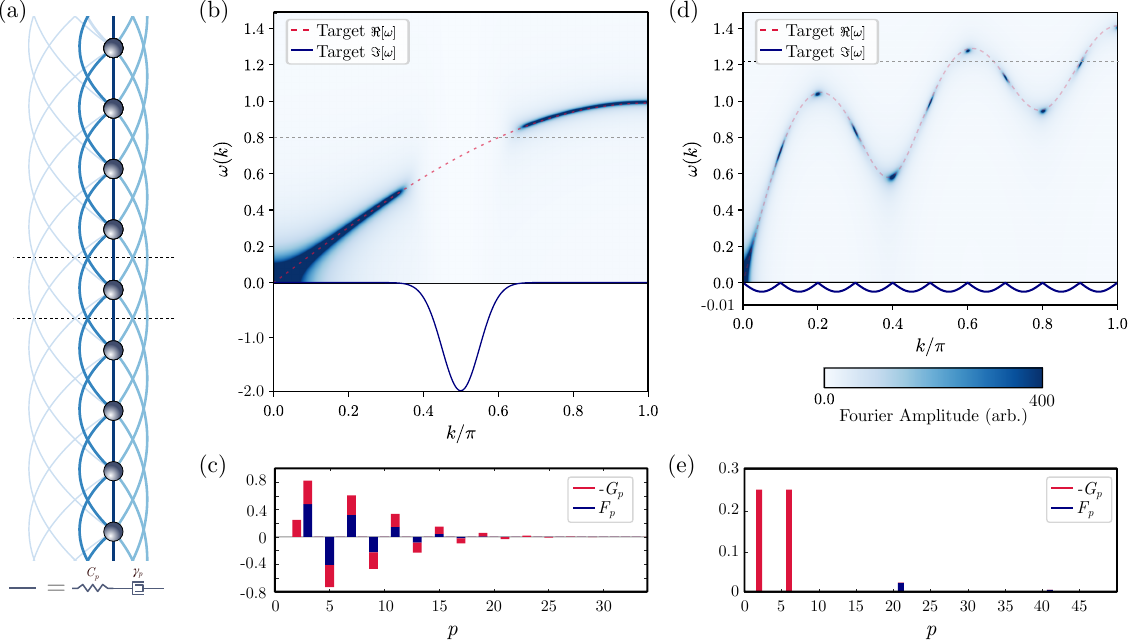}
    \caption{(a) Schematic of a BNN mass-spring chain, where inter-mass connections are modelled by a spring element (elastic restoring forces) and a dashpot element (dissipation or gain). (b) Fourier amplitude representation of an engineered dispersion curve, exhibiting a $k$-vector band gap. The target real and imaginary functions generating this curve are indicated by dashed red and solid blue lines, respectively. (c) Stiffness ($G_p$) and dissipative ($F_p$) Fourier coefficients derived from the inverse design protocol for the targets in (b). (d) depicts the Fourier amplitude representation of an exotic passive dispersion relation, designed via the positive definite convolution identity (Appendix~\ref{Ap: Positive Definite}), displaying a `dashed dispersion' - (e) shows the associated Fourier coefficients. The grey horizontal dashed lines in (b) and (d) highlight the excitation frequencies of the Green's functions solutions presented in Figure \ref{fig:greens}.}
    \label{fig: Hero}
\end{figure*}

Wave propagation, in structured materials or otherwise, is fundamentally governed by the dispersive properties of the media in which propagation occurs. There has been a concerted research effort to influence, even engineer, wave dispersion across almost all regimes \cite{Pendry2006, Wu2024, Li2018}. By design of a composite, underlying structure (typically subwavelength, periodic, and resonant), so-called metamaterials arise with unique, exotic, and even bespoke wave dispersion that can be utilised to control wave propagation for a variety of far reaching applications, such as the slowing and trapping of light \cite{Schulz2010, Jang2011, Tsakmakidis2007}, directional beam steering and lensing behaviours \cite{Aydin2007, Jia2018, Haxha2018, Craster2013}, enhanced acoustic absorption and noise reduction \cite{Gao2022, Mei2012, yangabsorp}, and energy harvesting \cite{Hu2021, DePontiHarvesting}, and seismic isolation \cite{seismicisolation, Colombi2016}, to name a few.

Lattices, in particular infinitely periodic structures, play a crucial role in the study of wave dynamics; their elegant analytical framework is often traced back to the seminal work of Brillouin \cite{Brillouin1953}. Recently there has been a renaissance in the ideas of Brillouin, incorporating non-local couplings that manifest as beyond-nearest-neighbour interactions through spatially extending beyond the unit cell \cite{Chen2021,Fleury2021, Wang2022, Moore2023, Martinez, Chaplainreconfig, Edge2025}. These couplings offer routes to engineering features in dispersion curves, for instance, Dirac points \cite{paul2025} or evanescent Bloch states \cite{Chen2024}, and enable the control of multiple (potentially arbitrarily many) regions of zero-group-velocity in the dispersion curves (band structures) of such crystals with applications in noise control \cite{Pelat2020}, energy harvesting \cite{Hu2021,Bilal2024}, and mode trapping \cite{Tian2020, Chaplainrainbowreflection}. It has been recently shown, via inverse design, that indeed any arbitrary dispersion curve on the real frequency-wavenumber ($\omega$-$k$) plane can be `drawn' \cite{Kazemi2023} in a lattice of masses and springs by leveraging the Fourier Series representation of the periodic band structure. The result being, with every additional Fourier component, additional non-local elements are required; for arbitrary curves this leads to a large number of beyond-nearest-neighbours, with possibly negative spring constants.    

Often, advances in such systems have been made using conventional, conservative mass-spring models i.e. dissipation is neglected. Important implications of this, particularly in the realm of elasticity have been highlighted \cite{Hussein2009,Hussein2010,Hussein2013,Bacquet2018}, where limits on harvesting applications have been derived in the context of `meta-damping', with extensions to Generalised Brillouin Zone theory considered \cite{Frazier2016}. Loss is an intrinsic property of materials, shaping the dynamical response of structures, and is often not possible, nor desirable, to be neglected \cite{Henriquez2017, Torrent2018,Acoublackhole}. Considered even less in such systems, is the inverse case; gain is seldom accounted for in simplistic mass-spring models, although with the advent of time-varying metamaterials \cite{Cho2020, Galiffi2021, Wen2022,Kim2024,Delory2024}, it is often a requirement and appears naturally in non-Hermitian systems.

In non-Hermitian systems gain and loss can give rise to unique wave-matter interactions \cite{Gu2021}, with solutions typically presenting as complex eigenvalues, due to instabilities and unconserved probabilities. Exceptions do exist to this, particularly in parity-time ($\mathcal{PT}$) symmetric systems which can stabilise and exhibit non-Hermitian behaviours \cite{Bender1998}. Demonstrating that, at critical points, the spontaneous breaking of $\mathcal{PT}$ symmetry can transition between stable systems with purely real eigenvalues and ones possessing complex eigenvalues \cite{Longhi2010}. Despite this, encompassing the principles of non-Hermitian behaviours can facilitate the emergence of new effects not previously possible in conventional systems, leading to unique wave-matter interactions and enhanced performance in a range of applications \cite{Ren2024}.

In this letter, we combine recent advances in beyond-nearest-neighbour (BNN) metamaterials with inverse design achieving arbitrary \textit{complex} dispersion relations through combining periodic mass-spring lattices with non-local interactions that include both gain and loss. Each spatial order of lattice connection is characterised by uniquely defined spring constants and damping terms (a schematic is illustrated in Figure \ref{fig: Hero}(a)). We then consider classes of sensible functions possessing only positive-definite Fourier coefficients, thus circumnavigating the requirement for gain or negative spring constants.

The result is a robust method that allows for a more nuanced control over the system's dispersion relation. We provide multiple examples that confirm the freedom of the design protocol and highlight the unique physics that can be achieved in such system, for example topical wavevector band gaps \cite{Chong2024, Kumar2018}. We present numerical solutions of the governing eigen-problem and of the system's Green's function, whilst providing commentary on the energy and group velocities in such systems and motivations towards physical realisation of such lattices.

\section{\label{Sec: Theory}Theory}

Our theory begins by considering damped oscillators in the canonical $1\mathrm{D}$ periodic mass-spring chain \cite{Frazier2016}. We begin with the case of monatomic masses connected via springs that may have arbitrary complex spring constants, i.e. they may contribute gain or loss to the system, and then expand this notation by incorporating $p$ beyond-nearest-neighbour (BNN) spatial connections, as visualised in Figure \ref{fig: Hero}(a). From Newton's second law of motion, we obtain a real-space equation of motion for the displacement ($\boldsymbol{U}_n$) of the $n^\text{th}$ point mass
\begin{align}
    \begin{split}
    m\ddot{\boldsymbol{U}}_n=\sum_{p=1}^{\infty} \Big[\gamma_p\left(\dot{\boldsymbol{U}}_{n+p}+\dot{\boldsymbol{U}}_{n-p} -2\dot{\boldsymbol{U}}_n\right)\\+ C_p\left(\boldsymbol{U}_{n+p}+\boldsymbol{U}_{n-p}-2\boldsymbol{U}_n\right)\Big],
    \end{split}
\end{align}
where $m$ denotes the mass, $C_p$ denotes the spring constant for each BNN order $p$ (i.e. connecting to the $p^{\text{th}}$ nearest neighbour), and $\gamma_p$ likewise represents the damping of the connection to the $p^{\rm th}$ nearest neighbour. Note we assume the physical size of the unit cell is unity ($a = 1$), and for harmonic excitations assume the convention ${\rm e}^{-{\rm i}\omega t}$ for e.g. displacement oscillations.

Due to lattice periodicity, we restrict our analysis to a single unit cell and employ Bloch's theorem \cite{Brillouin1953}, assuming the displacement in cell $n$ is related to that in the fundamental cell by $\boldsymbol{U}_n = \boldsymbol{U}_0{\rm e}^{{\rm i}{k}\cdot n}$, with $k$ the wavenumber along the chain. Thereupon we obtain the reciprocal space equation of motion
\begin{equation} 
    \underbrace{\left(\omega^2-{\rm i}\omega F(k) - G(k)\right)}_{R}\boldsymbol{U}=0, 
    \label{Eq: Reciprocal EOM}
\end{equation}
where
\begin{equation}
    F(k)=\sum_{p=1}^{\infty}2\beta_p\alpha_p,\qquad G(k)=-\sum_{p=1}^{\infty}\alpha_p\omega_{0,p}^2
\end{equation}
with $\beta_p = \frac{\gamma_p}{2m}$, $\omega_{0,p}^2=\frac{C_p}{m}$, and $\alpha_p = 2\left(\cos (kp)-1 \right)$. We refer to the bracketed term $R$ in \eqref{Eq: Reciprocal EOM} as the solvability condition, such that $R=0$ yields the lattice's dispersion relation.

The generalised form of \eqref{Eq: Reciprocal EOM}, permits the arbitrary exploration of the forward design problem, configuring lattices with unique combinations of nearest and beyond nearest neighbour spatial connections. Motivation is drawn from Kazemi \textit{et al.} \cite{Kazemi2023}, to investigate the \emph{inverse} design problem of general complex dispersion relations.

To address the inverse design problem, we first note that the dispersion curve is necessarily a repeating function, and so we can express any functional dependence of $\omega(k)$ as a Fourier series in $k$. Here the real functions $F(k)$ and $G(k)$ represent the dissipative and reactive behaviours of the lattice respectively, noting that $G(k)$ is a positive real function. The goal is to now find the BNN couplings $C_{p}$ and $\gamma_{p}$ that set desired real and imaginary parts of a given dispersion relation $\omega(k)$.   Clearly, $\Re[\omega (k)]$ prescribes the conventional band structure, whilst $\Im[\omega (k)]$ characterises the global non-conservative behaviours; a negative (positive) $\Im[\omega (k)]$ signifying damping (gain) within the system.

This inversion of the dispersion relation can be achieved through completing the square in \eqref{Eq: Reciprocal EOM}, 
\begin{align}
    \begin{split}
    \left(\omega(k)-\frac{{\rm i}F(k)}{2}\right)^2=G(k)-\frac{|F(k)|^2}{4},
    \end{split}
\end{align}%
enabling the identification of the two functions $F(k)$ and $G(k)$ in terms of the dispersion relation $\omega(k)$:
\begin{align}
    \begin{split}
     F(k) &= 2\Im[\omega(k)],\\
    G(k)&=\Re[\omega(k)]^2+\Im[\omega(k)]^2.\label{eq:inverse-design}
    \end{split}
\end{align}%
Here we note that $\Im[\omega (k)]$ and $\Re[\omega (k)]$ characterise the target dispersion functions, where $k$ is taken as a real parameter.  Equally the complex wavenumber/real frequency convention could be chosen.

The associated damping and stiffness Fourier coefficients appearing in \eqref{Eq: Reciprocal EOM} for each specified dispersion relation $\omega(k)$ are then found via a Fourier transform
\begin{align}
    \begin{split}
    F_p &= \int^\pi_{-\pi} F(k){\rm e}^{-{\rm i}kpa} {\rm d}k=\frac{\gamma_p}{m},\\
    G_p &= \int ^\pi _{-\pi} G(k) {\rm e}^{-{\rm i}kpa} {\rm d}k=-\frac{C_p}{m},
    \label{Eq: Fn, Gn Equations}
    \end{split}
\end{align}%
such that the damping and spring constants take real values. 

    The above Fourier expansion \eqref{Eq: Fn, Gn Equations} combined with the inverse design formulae \eqref{eq:inverse-design} enables us to find combinations of reactive and dissipative coupling constants for any complex dispersion relation $\omega(k)$. Whilst it is valuable to explore these theoretical limits, it is equally important to consider to the practical implications and the physical realisation of the structures under examination. Note that, if gain is required (negative Fourier coefficients $F_p$), a physical realisation necessitates active control. Similarly, negative spring constants ($C_p<0$) require more exotic inertial-resonant elements and are typically narrowband \cite{Bacquet2018}. Conversely, dispersion curves achievable without gain are theoretically obtainable through passive structures. Therefore, based on the sign of the Fourier coefficients defining the target dispersion relations, functions can be classified as either exotic passive (no gain) or exotic active (with gain). Motivated by the design of an exotic passive dispersion relation, we present a condition for designing target dispersion relations that ensures the inverse design protocol only yields real, positive-definite Fourier coefficients (see Appendix \ref{Ap: Positive Definite}).

\section{\label{Sec: Results}Results and Discussion}

\begin{figure}
    \centering
    \includegraphics[width=0.495\textwidth]{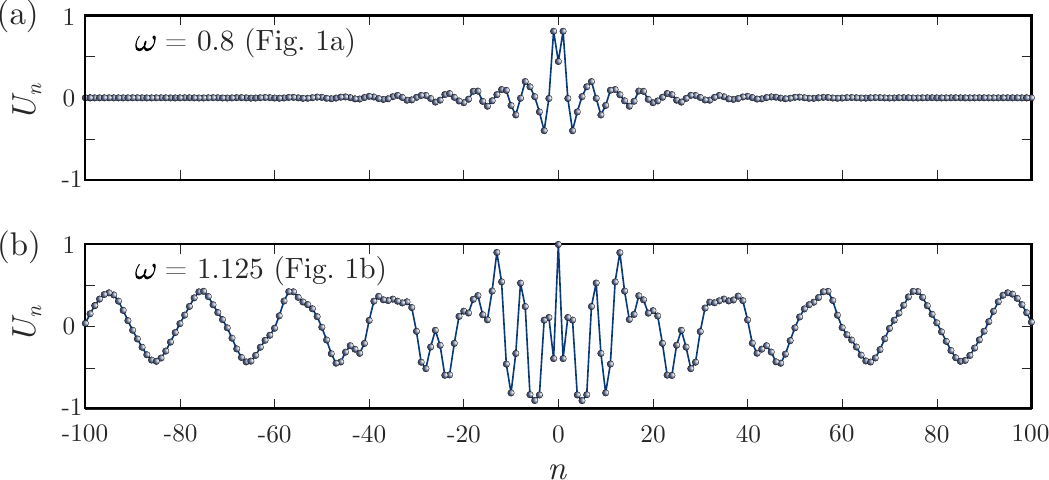}
    \caption{Green's function solutions (real point mass displacements) at select frequencies in the engineered dispersion curves presented in Figure \ref{fig: Hero}. (a) Solution driven at $\omega=0.8$ in dispersion curve in Figure \ref{fig: Hero}(b), in the apparent $k$-vector band gap. (b) Solution  driven at $\omega=1.125$, in dispersion curve in Figure \ref{fig: Hero}(d), which intersects the dispersion curve at three distinct $k$, two of which observe increased attenuation.}
    \label{fig:greens}
\end{figure}

The precise engineering of $\Im[\omega(k)]$ governs the attenuation characteristics within the designed dispersion profiles. When $\Im[\omega(k)]$ is tailored to exhibit a magnitude comparable to $\Re[\omega(k)]$, apparent band gap phenomena can be introduced into the complex dispersion relations, replicating behaviours observed in Hermitian systems. To highlight the power of this method, Figure \ref{fig: Hero}(b) presents an engineered system that exhibits a $k$-vector band gap in a \textit{real} fundamental phononic dispersion relation. Here $\Re[\omega(k)]$, is defined as the canonical nearest-neighbour monatomic mass-spring dispersion relation ($\Re[\omega(k)] \propto |\sin(ka)|$ \cite{Brillouin1953}), and $\Im[\omega(k)]$, is chosen as a Gaussian centred on $|k|=0.5\pi$ (given by \cite{note1}). To showcase the designed attenuation, our analysis is extended into the inhomogeneous regime, numerically solving the Green's Function $\left(\mathcal{F}^{-1}\left(R^{-1}\cdot \boldsymbol{f}\right)\right)$ for some applied point forcing $\boldsymbol{f}$, with $\mathcal{F}^{-1}$ denoting the inverse Fourier transform. Figure~\ref{fig:greens}(a) shows the Green's function at $\omega=0.8$, within the apparent $k$ band gap, illustrating evanescent wave behaviour. We perform the Green's function simulations, cycling through excitation frequency, and plot the absolute Fourier magnitude by the colour-scales in Fig.~\ref{fig: Hero}(b,d), where the apparent $k$-gaps are evident.

To generate the dispersion relation depicted in Figure \ref{fig: Hero}(b), a combination of both positive and negative Fourier coefficients, $F_p$ are necessary, as shown in Figure \ref{fig: Hero}(c). This implies that some of the lattice connections exhibit gain, i.e. by active control such as digital meta-atom approaches \cite{Cho2020}. Although interesting, practical considerations usually make gain difficult to implement.  This raises the question; can we restrict our inverse design formulae \eqref{eq:inverse-design} to yield only passive coupling parameters?

The answer to this question is yes; provided that careful consideration is given to the target function, $\omega(k)$. To eliminate the gain requirement, we introduce an additional design condition; we restrict all orders of Fourier coefficients, $F_p$ to positive definite values. We express $F(k)$ as the convolution of two $k$-dependent functions, $\left(h(k), s(k)\right)$ as outlined in Appendix~\ref{Ap: Positive Definite}. Imposing the condition $s(k)=h(-k)$, ensures real coefficients and taking the Fourier transform of $F(k)$ as detailed in \eqref{Eq: Fn, Gn Equations}, yields $F_n =|H|^2$, which is both real and positive. Here $H$ is the Fourier transform of $h(k)$.  As an example, the function $|\sin(\mu k)|$, satisfies this condition (proof in Appendix~\ref{Ap: Positive Definite}), and we use this in the application of our design method shown in Fig.~\ref{fig: Hero}(d).

\begin{figure}
    \centering
    \includegraphics[width=0.95\linewidth]{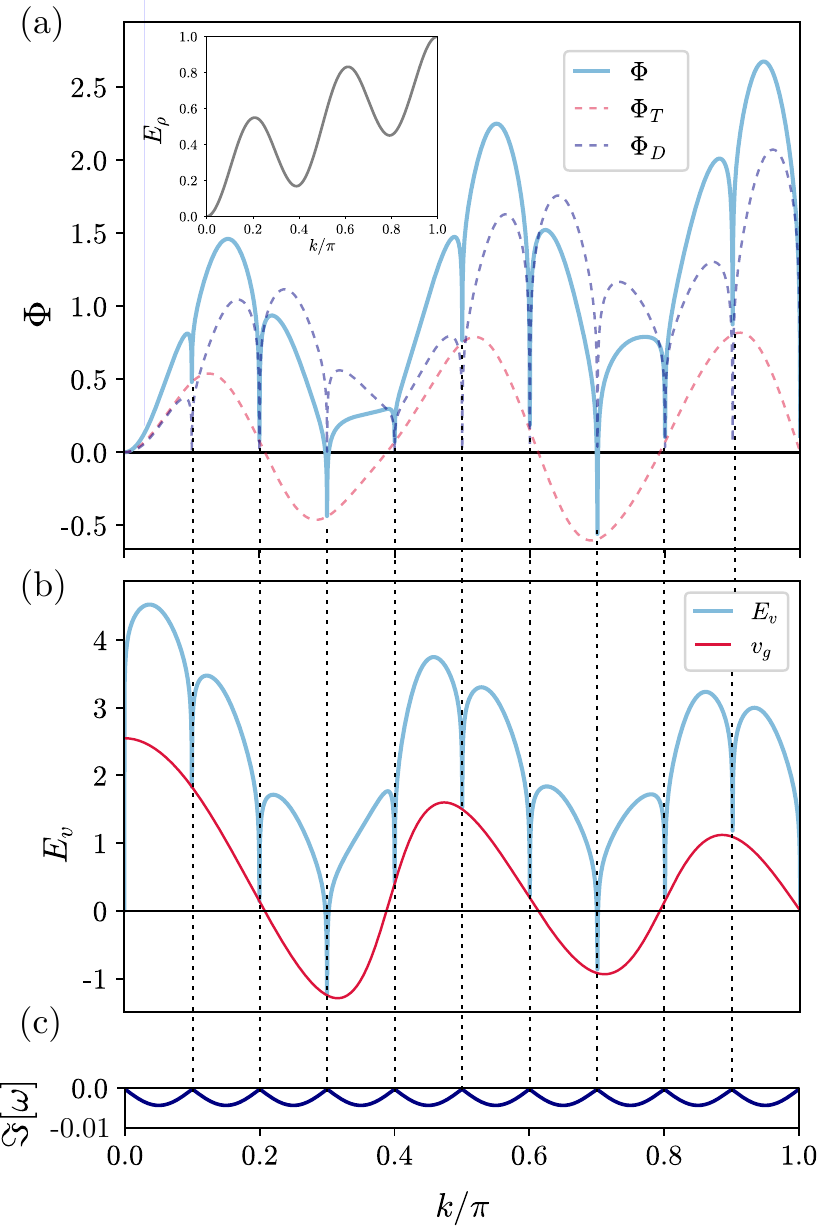}
    \caption{(a) Net energy flux arising from the kinetic energy of the mass for the systems presented in Figure \ref{fig: Hero}(d). $\Phi_T$ and $\Phi_D$, represent the transmitted and dissipated flux channels contributing to net power flow ($\Phi$), while the inset illustrates the energy density; the total kinetic and potential energy. (b) is the closely related energy velocity ($E_v$), which is the flux divided by the energy density, characterising the rate of energy flow. For comparison, the group velocity ($v_g$) of the equivalent undamped system ($\Re[\omega(k)]$ only) is presented. (C) is $\Im[\omega[(k)]$, outlining the designed attenuation.}
    \label{fig:energy-flux}
\end{figure}

Figure \ref{fig: Hero}(d) shows that the designed dispersion relation now incorporates a finite number of real, positive definite coefficients for both the dissipative, $F_n$, and reactive, $G_n$, response (see Figure \ref{fig: Hero}(e)). In this example, the real component, $\Re[\omega(k)]$, replicates the behaviour of a mass-spring lattice with first and fifth-order spatial connections, somewhat more complex than the recent `Roton-like' dispersion curves \cite{Chen2021},  while the imaginary component, $\Im[\omega(k)] = -|\sin(10k)|$, fulfils the aforementioned convolution identity. As shown in Figure \ref{fig: Hero}(d), increased attenuation is observed in line with the maximal points of $\Im[\omega(k)]$, creating a dashed-like appearance (Fourier amplitude in Fig.~\ref{fig: Hero}(d)). This is further demonstrated in Figure \ref{fig:greens}(b), which displays the Green's function at a driving frequency ($\omega=1.125$) that intersects three $k$ solutions, two of which observe increased attenuation. The decay envelopes of the decay modes are localised near $n=0$, while a singular driven mode dominates outside this region.

We now investigate the energy transport within the structure corresponding to the dispersion shown in Figure \ref{fig: Hero}(d).  One approach to this would be to consider the elastic Poynting vector and group velocity, however, in this non-Hermitian regime we must be careful in applying conventional wave analysis techniques. Dissipative effects engenders a complex reshaping of a wave packet during propagation, meaning that we must be careful in inferring information from the group velocity, something known well in both the electromagnetic and elastic literature \cite{bernard1999energy,chen2010group}. Here we explore energy transport throughout the lattice via the energy flux and the closely related energy velocity, which characterises the rate at which energy propagates within a lattice.  Specifically we consider the rate of loss of kinetic energy in a given unit cell, finding the contributions due to propagation and dissipation. Appendix \ref{Ap:Elastodynamic Poyntings Theorem} contains the derivation of energy flux and velocity following the approach of Brillouin. 

We delineate between the two constituent contributions to the total energy flux, namely: the dissipated energy flux ($\Phi_D$), representing energy lost within the system, and the transmitted energy flux ($\Phi_T$), which quantifies the energy transfer between adjacent unit cells such that the total energy flux (rate of loss of kinetic energy) is $\Phi = \Phi_T + \Phi_D$.  We show the derivation (generalised from Brillouin \cite{Brillouin1953} to included dissipation and BNNs) in Appendix~\ref{Ap: energy flux}:
\begin{align}
\begin{split}
    \Phi &= \sum_p^P \Bigg(\frac{U_0^2 C_p\Re(\omega)p}{2}\sin(kp)\\
    +&p\left(\frac{U_0^2C_p\Im(\omega)}{2}+\frac{U_0^2 \gamma_p |\omega|^2}{2}\right) \left(1-\cos\left(kp\right)\right)\Bigg),
\end{split}
\end{align}
where the first line gives the contribution of $\Phi_T$ with the second giving $\Phi_D$, respectively.
Graphically shown in Figure \ref{fig:energy-flux}(a), $\Phi_D$ converges to zero in alignment with the tailored dissipation profile (Figure \ref{fig:energy-flux}(c)), reflecting the critical role of $\Im[\omega(k)]$. Moreover, the dissipated energy flux remains strictly positive ($\Phi_D>0~\forall~k\geq 0$), indicating the absence of gain mechanisms within the system, reinforcing the exotic passive condition arising from the convolution identity. Complimentary to this, the transmitted energy flux exhibits features akin to those of the equivalent un-damped system, revealing localised regions where $\Phi_T=0$; these regions correspond to turnover points in $\Re[\omega(k)]$. Equally significant is the development of negative regions in the transmitted energy flux, a consequence of the competition between power channels, which drives a system response exhibiting apparent retrograde energy flow \cite{Edge2025}.

Figure \ref{fig:energy-flux}(b) plots the closely related energy velocity, defined as the energy flux divided by the energy density (Fig.~\ref{fig:energy-flux}(a), inset) -- this is required to be analysed as the notion of group velocity loses significant meaning in non-conservative systems \cite{chen2010group}. For completeness we show the group velocity of $\Re[\omega(k)]$; not shown is $v_g(k=0)=0$ - this discrepancy arises from a discontinuity in $v_g$ at $k=0$, which separates the positive $(k\geq 0)$ and negative $(k<0)$ regions of the first Brillouin zone. The distinction between $E_v$ and $v_g$ elucidates the expected challenges in characterising energy propagation via the group velocity in dissipative media. Yet, there is an expected commonality when dissipation becomes negligible ($\Im[\omega(k)]\rightarrow 0$), with $E_v \rightarrow v_g$, revealing their analogous behaviour in conservative systems.

Before concluding, we note here that with this powerful theoretical freedom there are some challenges to highlight; high-gradients in target dispersion relations are sensitive to Gibbs phenomena \cite{Gibbs1998}, prompting the exploration of continuous, smooth functions to mitigate this. Furthermore the designed dispersion relations are constrained by the condition, $\omega(k=0)=0$, arising from the recurring $\alpha_p$ term in the previously defined equation of motion \eqref{Eq: Reciprocal EOM}, which aligns with the physical motivation presented in \cite{Kazemi2023}. However, it's important to recognise that this condition is not universally applicable, as demonstrated in systems with zero-frequency band gaps, where the behaviour at $k=0$ can be fundamentally different \cite{Achaoui2017, Yang2021}.

 \section{Conclusions}

Dispersion engineering offers a compelling avenue for advancing technologies across a diverse range of fields such as noise reduction, seismic isolation, sensing, and even energy harvesting \cite{Li2018, Colombi2017, Chaplainreconfig}. We have given a new, versatile inverse design protocol where both the real and imaginary parts of any dispersion relation can be related to a fixed set of BNN coupling constants.  We have applied this method to explore a set of engineered dispersion relations, which includes the formation of apparent $k$-vector band gaps, and `dashed' dispersions, where only narrow ranges of $k$ values exhibit lossless propagation. Extending the inverse design of lattice couplings to encompass non-Hermitian behaviour thus provides a powerful framework for tailoring wave propagation characteristics.

\begin{acknowledgements} 
\noindent R.G.E. and T.A.S. acknowledge the financial support of Defence Science and Technology Laboratory (Dstl) through grants, DSTL0000022047, DSTLXR1000154754 and AGR 0117701. S.A.R.H. and G.J.C. acknowledge the financial support by the EPSRC (grant no EP/Y015673/1). All data created during this research are available upon reasonable request to the corresponding author. ‘For the purpose of open access, the author has applied a ‘Creative Commons Attribution (CC BY) licence to any Author Accepted Manuscript version arising from this submission’.
\end{acknowledgements}


%

\appendix
\section{Real Positive Definite Fourier Coefficients}
\label{Ap: Positive Definite}
Here we present the methodology for ensuring positive definite Fourier coefficients (and thereby leading to passive structures) in terms of the function $F(k)$ -- this is a general condition which can also apply to $G(k)$.

Assuming that $F(k)$ can be expressed as the convolution of two $k$ dependent functions $h(k)$ and $s(k)$ such that the convolution is defined by, 
\begin{equation}
    F(k) = (h*s)(k) = \int h(t)s(k-t)dt,
\end{equation}
we establish a framework to impose the real and positive definite Fourier coefficient on the design constraints. As detailed by \eqref{Eq: Fn, Gn Equations}, $F_p$ is expressed as the Fourier transform of $F(k)$, 
\begin{equation}
    F_p = \int F(k)e^{ikpa}dk.
\end{equation}
Thus, applying the Fourier transform and utilising the convolution theorem 
\begin{equation}
    F_p= \mathcal{F}(F(k)) = \mathcal{F}\left((h*s)(k)\right) = \mathcal{F}(h)\mathcal{F}(s), 
    \label{eq: Convolution identity}
\end{equation}
i.e. the Fourier transform of a convolution is the product of the Fourier transformed functions, informs as to the nature of the $F_p$ coefficients. To ensure that $F_p$ is real, $s(k)=h(-k)$, is imposed which when substituted into \eqref{eq: Convolution identity} which returns 
\begin{align}
\begin{split}
    F_p =&~\mathcal{F}\left((h*h(-k))(k)\right)= \mathcal{F}((h*h^\dagger)(k))\\
    =&~\mathcal{F}(h(k))\mathcal{F}(h^\dagger(k)) = HH^\dagger = |H|^2,
\end{split}
\end{align}
where the $\dagger$ indicates the complex conjugate and $H$ refers to the Fourier transform of $h(k)$, $H = \mathcal{F}[h(k)]$. As highlighted by $|H|^2$, if the function $F(k)$ satisfies the convolution identity such that $s(k)=h(-k)$, the resulting Fourier coefficient will always be real and positive definite. Thus through the strategic selection of input functions for the inverse design protocol, a structure can be designed as either exotic active or exotic passive. The function
\begin{equation}
    F(k) = -\lambda|\sin(\mu k)|
    \label{eq: mod sin}
\end{equation}
satisfies the convolution identity of the exotic passive classification and offers the advantage of requiring a minimal orders of Fourier coefficients to effectively resolve. Here, $\mu~\in~\mathbb{{N}^+}$ dictates the period of the designed function, while $\lambda$ is a positive coefficient defining the amplitude. 

Substituting \eqref{eq: mod sin} into \eqref{Eq: Fn, Gn Equations} gives,
\begin{equation}
    F_p = -\frac{2\lambda}{\pi} \int_{0}^{\frac{\pi}{\mu}} \sin(\mu k)e^{-2ikp} dk, 
\end{equation}
which we rewrite in terms of complex exponentials,
\begin{equation}
    F_p = -\frac{2\mu\lambda}{\pi} \int_{0}^{\frac{\pi}{\mu}} \frac{e^{-2i\mu kp}}{2i}\left(e^{i\mu k}-e^{-i\mu k}\right) dk,
\end{equation}
where $p \in \mathbb{N}^+$ is the range of spatial connections. Subsequently performing the integral and applying the limits yields
\begin{align}
    \begin{split}
    F_p =  -\frac{\mu\lambda}{\pi}\Big(\frac{e^{-i\pi(2p-1)}}{\mu(2p-1)}-\frac{e^{-i\pi(2p+1)}}{\mu(2p+1)} \\\\ -\frac{1}{\mu(2p-\mu)}+\frac{1}{\mu(2p+\mu)} \Big),
    \end{split}
\end{align}
simplifying to
\begin{align}
    \begin{split}
    F_p =  -\frac{\mu\lambda}{\pi}\Big(\frac{\cos(\pi(2p-1))}{\mu(2p-1)}&-\frac{\cos(\pi(2p+1))}{\mu(2p+1)} \\\\ -\frac{1}{\mu(2p-1)}&+\frac{1}{\mu(2p+1)}\Big),
    \end{split}
\end{align}
due to the orthogonality of sines and cosines. Further simplifying $F_p$ via
\begin{equation}
    \cos(\pi(2p\pm1)) = (-1)^{2p\pm1}= -1,
\end{equation}
we arrive at
\begin{align}
\begin{split}
    F_p =  -\frac{2\lambda}{\pi}\left(\frac{1}{(1-2p)}+\frac{1}{(1+2p)} \right)
    = \frac{4\lambda}{\pi(4p^2-1)}
\end{split}
\end{align}
which satisfies $F_p > 0 ~\forall~ p$. For completeness the functions $h(k)$ and $s(k)$ of the $F(k)$ convolution identity can be determined via 
\begin{align}
\begin{split}
    h(k) = \mathcal{F}^{-1}\left(\sqrt{F_p}\right),\\
    s(k) = \mathcal{F}^{-1}\left(-\sqrt{F_p}\right).
\end{split}
\end{align}

We note that the Gaussian loss function will never incorporate strictly positive-definite Fourier coefficients due to the finite bounds imposed by the Brillouin zone width. Thus, application of the convolution identity, yields some error function representation:

\begin{align}
\begin{split}
    (h*s)(k) =& \int_{-\frac{\pi}{a}}^{\frac{\pi}{a}} h(u)s(k-u)du\\
    =& ~\xi(k)\cdot \Big[\text{Erf}\big(\tau \left(u-\chi (k)\right)\big)\Big]_{\frac{-\pi}{a}}^{\frac{\pi}{a}}.
\end{split}
\end{align}
Here $\tau$ represents a $k$ independent coefficient associated with the standard deviations of the convolved Gaussians, while $\xi(k)$ and $\chi(k)$ are some unspecified $k$ dependent functions, with no guarantee of being positive definite.

\section{\label{Ap:Elastodynamic Poyntings Theorem} Energy Flux and Velocity}

An elastic body's behaviour can be described via generalised Hooke's law,
\begin{equation}
    \sigma _{ij}=C_{ijkl}\epsilon_{kl},
\end{equation}
where $\sigma _{ij}$ is the stress tensor, $C_{ijkl}$ is the elasticity tensor, and $\epsilon_{kl}$ is the strain tensor. The elastodynamic Poynting's theorem then provides a statement of the energy conservation within that system, 
\begin{align}\begin{split}
        \frac{\partial U}{\partial t}+ \nabla\cdot\boldsymbol{S_i}+P_d=P_{ext}.
\end{split}\end{align}
Here, $P_d$ is the dissipated power, $P_{ext}$ is the external power, and $U$ is the total kinetic and potential energy stored per unit volume, such that $\frac{\partial U}{\partial t}$ is the rate of change of the energy density i.e. $-\rm{i}\omega$$U$. Additionally, $\nabla\cdot\boldsymbol{S_i}$ is the divergence of the Poynting vector $\left(\boldsymbol{S_i}=-(\boldsymbol{\sigma}_{ij} \cdot \boldsymbol{v}_j)\right)$, describing the energy flow out of a unit volume. 
Throughout this letter we consider a non-Hermitian inverse design protocol, where the examined structures exhibit dissipative behaviours. As a result of this any wave analysis via the group velocity is restricted as attenuation alters the nature of a wave's propagation resulting in undefined wave packets. Consequently, we examine wave propagation throughout the lattice via the energy flux, particularly using the complex time-averaged energy flux due to the time harmonicity of the analysis, 
\begin{equation}
    \langle\boldsymbol{S}_i\rangle = -\frac{1}{2}\boldsymbol{\sigma}_{ij} \boldsymbol{v}_j^*= -\frac{1}{2}\Re(\boldsymbol{\sigma}_{ij} \boldsymbol{v}_j^*) + \frac{-\rm{i}}{2}\Im(\boldsymbol{\sigma}_{ij} \boldsymbol{v}_j^*).
\end{equation}

In evaluating the complex time-averaged energy flux, two quantities are obtained. The first being the reactive power ($-\frac{i}{2}\Im(\boldsymbol{\sigma}_{ij} \boldsymbol{v}_j^*)$), which describes the energy oscillating between the spring and the mass, due to a phase difference between the mass displacement and the forcing. The other being the active power ($-\frac{1}{2}\Re(\boldsymbol{\sigma}_{ij} \boldsymbol{v}_j^*)$), which describes useful work done on the system, either by dissipation, or via power transmitted to the neighbouring unit cell. This will be the quantity we focus our analysis on. Evidently, in the case of masses and springs the above notation simplifies significantly to the usual, undergraduate form of linear Hooke's law.

\subsection{\label{Ap: energy flux}Energy Flux}
The nearest neighbour time-averaged energy flux (i.e. we omit the sum over BNNs), can be expressed in terms of the displacement vector $U_n$, and the dissipative and restorative forces,

\begin{equation}
    \Phi = -\frac{1}{2}\Re \left(\left(C(U_{n+1}-U_{n})-\mathrm{i}\omega \gamma ({U}_{n+1}-{U}_{n})\right) \cdot \left(-\mathrm{i} \omega {U}_n\right)^{\dagger}\right),
\end{equation}
where all symbols are defined as in the main body of text. However, owing to the fact we are considering dissipative lattices with complex frequency profiles, that experience temporal dissipation, careful attention is paid to both the sign of $k$ and $\omega$ to avoid violating causality. Thus the energy flux is calculated for $k\geq0$ in the first Brillouin zone, where the positive propagation direction is defined from $0\rightarrow \infty$. Next by utilising the Bloch phase,
\begin{align}
\begin{split}
    \Phi = -\frac{1}{2}\Re \Bigg(\Big(C&\left(e^{\mathrm{i}k}-1\right)\\
    -& \rm{i} \omega \gamma \left(e^{\rm{i}k}-1\right)\Big)U_n \cdot (-\rm{i}\omega \it{U}_n)^{\dagger}\Bigg),
\end{split}
\end{align}
and employing Euler's identity, we get
\begin{align}
\begin{split}
    \Phi = -\frac{U_0^2}{2}\Re \Big(\rm{i}&\omega^\dagger C\left(\cos(k)+\rm{i}\sin(k)-1\right)\\
    +&|\omega|^2 \gamma \left(\cos(k)+\rm{i}\sin(k)-1\right)\Big),
\end{split}
\end{align}
enabling the real part of the energy flux to be determined:
\begin{align}
\begin{split}
    \Phi =& \frac{U_0^2 C\Re(\omega)}{2}\sin(k)\\ 
    &+\left(\frac{U_0^2C\Im(\omega)}{2}+\frac{U_0^2 \gamma |\omega|^2}{2}\right) \left(1-\cos(k)\right).
\end{split}
\end{align}
Here the first $\sin(k)$ dependent term correlates with the transmitted energy flux passed to the neighbouring unit cell. In contrast, terms involving $(\cos(k)-1)$ correspond to the dissipated power within the first Brillouin zone, exhibiting spatial invariance as indicated by the even parity associated with  $\cos(k)$. The generalised energy flux notation incorporating BNN spatial connections is straightforwardly expressed as we show in the main text.

\subsection{\label{Ap: energy velocity}Energy Velocity}

A quantity closely linked to energy flux is the energy velocity, which characterises the rate at which energy propagates within a lattice. It is defined as the ratio of energy flux to energy density, 
\begin{equation}
    E_\nu = \frac{\Phi}{E_{\rho}},
\end{equation}
 where, $E_{\rho}$, the energy density is the sum of the kinetic energy $(T)$ and the potential energy $(V)$.

\begin{align}
\begin{split}
    E_{\rho} =& \frac{1}{2}mv^2-\frac{Cx^2}{2}\\\\
    =& \frac{1}{a}\Re \left(\frac{1}{2}m \dot{U}_n^2\right)+\frac{1}{a}\Re\left(\frac{C}{2}(U_n-U_{n+1})^2\right)
\end{split}
\end{align}
As with the energy flux, we consider the real time averaged energy density, which after some mathematical manipulation simplifies to,

\begin{equation}
    E_{\rho} = \frac{mU_0^2|\omega|^2}{4a}+\frac{CU_0^2}{2a}(1-\cos(k)).
\end{equation}
whose BNN generalised equivalent is:
\begin{equation}
    E_{\rho}^{BNN} = \sum_p^P \left(\frac{mU_0^2|\omega|^2}{4a}+\frac{C_pU_0^2}{2a}(1-\cos(kp))\right).
\end{equation}

\end{document}